\documentclass[prd,nofootinbib,longbibliography,twocolumn,preprintnumbers]{revtex4-1}

\usepackage[mathscr]{eucal}
\usepackage[latin1]{inputenc}
\usepackage{amsmath}
\usepackage{amsfonts}
\usepackage{amssymb}
\usepackage[usenames,dvipsnames]{color}
\usepackage[pdftex]{graphicx}
\usepackage[raggedleft]{sidecap}

\usepackage{subfig}
\usepackage{graphicx}
\usepackage{sidecap}
\usepackage{hyperref}

\newcommand{\C}{\mathbb{C}}

\def\tr{{\rm Tr}}

\newcommand{\f}{\frac}

\newcommand{\fracs}[2]{{\scriptstyle\frac{#1}{#2}}} 

\usepackage{enumerate}

\usepackage{colordvi}
\newcounter{mnotecount}[section]

\newcommand{\comment}[1]{}
\def\f{\frac}

\def\d{\textrm{d}}

\usepackage{enumerate}

\newcommand{\be}{\nopagebreak[3]\begin{equation}}
\newcommand{\ee}{\end{equation}}
\newcommand{\ba}{\nopagebreak[3]\begin{eqnarray}}
\newcommand{\ea}{\end{eqnarray}}
\newcommand{\nn}{\nonumber \\ }

\newcommand{\bra}[1]{\ensuremath{\langle#1|}}
\newcommand{\ket}[1]{\ensuremath{|#1\rangle}}

\def\ini{\textrm{i}}
\def\fin{\textrm{f}}

\definecolor{celeste}{rgb}{.23,.5,.7}
\newcommand{\Feyn}[1]{#1\kern-0.65em/}
\def\be{\begin{eqnarray}}
\def\ee{\end{eqnarray}}
\def\bc{\begin{center}}
\def\ec{\end{center}}

\def\vol{{\mathtt{v}}}

\begin{document}

\title{Many-nodes/many-links spinfoam: the homogeneous and isotropic case}
     \author{Francesca Vidotto}
 \email{vidotto@cpt.univ-mrs.fr\\[1em] 
{${} \!\!\!\!\!\! \heartsuit$}\,Unit\'e mixte de recherche (UMR 6207) du CNRS et des Universit\'es de Provence (Aix-Marseille I), de la M\'editerran\'ee (Aix-Marseille II) et du Sud (Toulon-Var); laboratoire affili\'e \`a la FRUMAM (FR 2291).
 }        
 
     \affiliation{\vspace{2mm}Centre de Physique Th\'eorique de Luminy${\,}^\heartsuit\!\!$, 
     Case 907, F-13288 Marseille, EU;\\
     Dipartimento di Fisica Nucleare e Teorica,
        Universit\`a degli Studi di Pavia, and\\  Istituto Nazionale
        di Fisica Nucleare, Sezione di Pavia, via A. Bassi 6,
        I-27100 Pavia, EU.}
 

\begin{abstract}                 \vskip2mm

\noindent 
I compute the Lorentzian EPRL/FK/KKL spinfoam vertex amplitude at the first order for regular graphs, with an arbitrary number of links and nodes, and coherent states peaked on a homogeneous and isotropic geometry. This form of the amplitude can be applied for example to a dipole with an arbitrary number of links or to the 4-simplex given by the complete graph on 5 nodes.  
All the resulting amplitudes have the same support, independently of the graph used, in the large $j$ (large volume) limit. This implies that they all yield the Friedmann equation: I show this in the presence of the cosmological constant.  This result indicates that in the semiclassical limit quantum corrections in spinfoam cosmology do not come from just refining the graph, but rather from relaxing the large $j$ limit.
\vspace{20mm}
\end{abstract}

\maketitle
\section{\!\!\!\!\!\!ntroduction}

The covariant (path-integral) approach to quantum cosmology 
consists in the computation of transition amplitudes between two
quantum states that describe the geometry of the universe 
 This can be done in particular in
minisuperspace models 
, where the infinite number of degrees of
freedom of General Relativity is truncated to a finite number.

In Loop Quantum Gravity all these ingredients are well defined: the
path integral is formulated in the spinfoam formalism, where the sum
is over possible geometries, the states are spinnetwork states from
which one can construct coherent states peaked on a given geometry,
and finally the truncation on a graph of the theory provide a natural
way to obtain a finite number of degrees of freedom. (For an
introduction, see for example \cite{Rovelli:2011eq}.)

  \begin{SCfigure}[2][b]
  \includegraphics[width=6em]{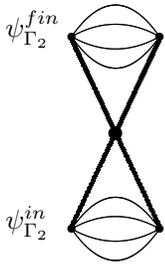}\hspace{4mm}
\caption{Transition amplitude between two states defined on a ``dipole" graph. I consider only the first order in the vertex expansion, i.e. there is only one vertex in the bulk (spinfoam edges are drawn with thicker lines).}
\end{SCfigure}

\newpage

The EPRL/FK/KKL spinfoam amplitude  \cite{Engle:2007uq,Livine:2007vk,Engle:2007qf,Freidel:2007py,Engle:2007wy,Kaminski:2009fm} has been evaluated in the Euclidean framework  for a homogeneous isotropic geometry on a particularly simple graph, and given a tentative cosmological interpretation \cite{Bianchi:2010zs}, opening up the possibility of studying quantum cosmology directly from the spinfoam formalism.

Various questions remain to be addressed, however, in order to make such \emph{spinfoam cosmology} viable \cite{Bianchi:2010zs,Bianchi:2011ym,Vidotto:2010kw,Roken:2010vp,Hellmann:2011jn}.  First, the result must be extended to the Lorentzian context.   Second,
spinfoam cosmology is based on the idea of approximating the spinfoam sum with its value on simple graphs and two-complexes. Is this expansion viable? The graph used so far is the \emph{dipole graph}  \cite{Rovelli:2008ys,Battisti:2010kl,Battisti:2010uq,Borja:2011di} given by $N=2$ nodes of equal valency (or degree) $d=4$. This graph has a nice geometrical interpretation, being the simpler graph that can be associated to the triangulation of a 3-sphere. What happens if we use a different graph?  In general, the choice of the graph determines the number of degrees of freedom taken into account; in the semiclassical limit of a homogeneous isotropic configuration these should not matter.  How is the spinfoam cosmology transition amplitude modified by using a different graph, namely adding more links and/or more nodes?

I address some of these issues: I generalize spinfoam cosmology to an arbitrary regular graph with many nodes and many links, and to the Lorentzian framework.   
I show that the semiclassical behavior of the model is the same as in the Euclidean and it is independent from the graph chosen. The transition amplitude turns out to be modified just by a global factor, in a way much similar to what happens for Regge calculus \cite{Collins:1973mh}.

This result supports the idea that the graph expansion is consistent in spinfoam cosmology and indicates that quantum correction to the Friedmann dynamics in spinfoam cosmology are not given by more complicated graphs, but rather to subleading terms in the semiclassical, large volume, limit.   
I refer only to graphs on the boundary, while I do not address in this paper the issue of refining the two-complex in the bulk (for a discussion about this see \cite{Rovelli:2011tt}).

In this paper I discuss a covariant quantum cosmology defined from the full quantum gravity theory in the spinfoam formalism. A different approach has been recently explored  \cite{Ashtekar:2009dn,Rovelli:2009kx,Ashtekar:2010ve,Campiglia:2010jw,Ashtekar:2010fk,Henderson:2010qd,Calcagni:2010ad}
starting from the Hamiltonian constraint of Loop Quantum Cosmology and defining a path integral, that mimics the expansion in spinfoam. The two approach should hopefully converge.

\vspace{1em}

In the next section I compute the Lorentzian EPRL/FK/KKL transition amplitude in the homogeneous and isotropic case for a general abstract graph. In Sec.\,\ref{secos}  I introduce the cosmological constant in order to study the resulting Friedmann equation.  Finally, in Sec.\,\ref{examples},  I briefly discuss two special cases of this amplitude: the dipole with many links and the 4-simplex-boundary, given by the complete graph on five nodes $\Gamma_5$.

 \section{Transition amplitude}
 
The EPRL/FK/KKL spinfoam amplitude has the form 
 \be\label{zeta}
   Z_{\cal C}=\sum_{j_f,\vol_e} \prod_f (2j_f+1) \prod_v A_v(j_f,\vol_e). 
   \label{old}
\ee
where $A_v(j_f,\vol_e)=\langle j_f,\vol_e| A_v\rangle$ is the vertex amplitude in the spin network basis. (See \cite{Rovelli:2011eq} for an introduction to this formalism and full definitions.)  

I use the coherent states
\cite{Bianchi:2009ky,Bianchi:2010ys,Bianchi:2010zr}
\be
\psi_{H_\ell}(U_\ell) = \int_{SU(2)^N}\! dg_n 
  ~     \prod_{l\in \Gamma} K_{t}(\, g_{s(\ell)}\, U_\ell \, g^{-1}_{t(\ell)} \ H_\ell ^{-1})
 \label{coherent}
 \ee
as boundary states for the transition amplitudes. 
They are defined by an integral on $SU(2)$, so that the states are gauge invariant, and  by the \emph{heat kernel} $K_t$ on $SU(2)$ $\big(U_\ell\in SU(2)\big)$, analytically continued to $SL(2,\C)$. This is a function concentrated on the origin of the group, with a spread%
\footnote{The parameter $t$, called the Heat-kernel time, is taken here with the dimension of an action. The coherent states became classical for small values of this parameter.}
 of order~$1/t$ in $j$.
These states are labelled by one element  $H_\ell\in SL(2,\C)$ for each link. This can be written as %
\be\label{acca}
 H_\ell= 
 D^{\scriptscriptstyle(j)}(R_{\vec n_{s(\ell)}})\ 
e^{-iz_\ell\f{\sigma_3}2}\ 
D^{\scriptscriptstyle(j)}(R_{\vec n_{t(\ell)}}^{-1})~.
 \ee
where $R_{\vec n}\in SU(2)$ is some fixed choice of rotation matrix that rotates the unit vector pointing in the $(0,0,1)$ direction into the unit vector $\vec {n}$, and $D^{(j)}(R_{\vec n_s})$ is its representation $j$. The geometrical interpretation is the following \cite{Freidel:2010uq,Magliaro:2010vn}.  The two vectors $\vec n_s$ and $\vec n_t$ represent the normals to the face $\ell$, in the two polyhedra bounded by this face. The complex number $z_\ell$ codes the intrinsic and the extrinsic geometry at the face. More precisely the imaginary part of $z_\ell$ is proportional to the area of the face of the triangulation dual to the link $\ell$. The real part of $z_\ell$ is determined by the holonomy of the Ashtekar connection along the link \cite{Rovelli:2010km}. 

I focus on the evaluation of the single vertex amplitude $A_v$. When evaluated in the (holomorphic) basis of the coherent states \eqref{coherent}, the amplitude is
\be
W(H_\ell)=\langle A_v | \psi_{H_\ell}  \rangle
\ee
this can be written as \cite{Pereira:2007nh,Engle:2008ev,Barrett:2009bs,Bianchi:2011ta}:
\be\!\!\!\!\!\!
W(H_{\ell})=  \int_{SL(2,\C)}\ \prod_{n=1}^{N\!-\!1}\! \d G_n \;
\prod_{\ell=1}^ {L}\;     P_t(H_{\ell}, G_\ell)~
\ee
where
\begin{multline}
\label{pirippipipi}
P_t(H_\ell,G_\ell) = 
\sum_{j_\ell} {\scriptstyle (2j_\ell+1)}e^{\scriptscriptstyle\!-\!2t\hbar j_\ell(j_\ell+1)} 
\\ \times 
\ \tr\!
\left[\! 
D^{\scriptscriptstyle(j_\ell)}\!(H_\ell)
Y^\dagger  D^{\scriptscriptstyle(\gamma j_\ell, j_\ell)}\!(G_\ell) Y
\! \right]. 
\end{multline}
$
D^{\scriptscriptstyle(j)}(H_\ell)$ is simply
$
D^{\scriptscriptstyle(j)}(R_{\vec n_{s(\ell)}})\ 
D^{\scriptscriptstyle(j)}(e^{-iz_\ell\f{\sigma_3}2})\ 
D^{\scriptscriptstyle(j)}(R_{\vec n_{t(\ell)}}^{-1})~.
$
$G_\ell=G_{s(\ell)}G_{t(\ell)}^{-1}$ is the product of the $SL(2,\C)$ group elements at the source and target nodes, extremals of each oriented link $\ell$, and $D^{\scriptscriptstyle(\gamma j_\ell, j_\ell)}\!(G_\ell)$ is its representation matrix.
Finally,
$Y$ is a map from the representation $(j)$ of $SU(2)$ to the representation $(\gamma j_\ell, j_\ell)$ of $SL(2,\C)$. (We denote with $\gamma$ the Barbero-Immirzi parameter.)

I want to evaluate this expression in the homogeneous and isotropic case. 
This corresponds to restricting the study to regular graphs \cite{Rovelli:2010vn}, so that the distribution of the degrees 
of the nodes is uniform (all the nodes have the same valence).  The requirement of homogeneity and isotropy
fixes $\vec n_s, \vec n_t$ as the normals to the faces of the geometrically regular cellular decomposition dual to 
the graph, and implies that all the $z_\ell$ elements in $H_\ell$ are equal: $z_\ell=z$.  Furthermore, on a
homogeneous isotropic space the real part of $z$ is the sum of two terms \cite{Magliaro:2010qz}
\ba\label{Im2}
\textrm{Re}\,  z = \theta (\gamma K + \Gamma)
~,
\ea  
where $K$  and $ \Gamma $ are the scalar coefficients of respectively the extrinsic curvature  and the spin connection, that enter in the definition of the Ashtekar-Barbero connection written in the homogeneous gauge. On a compact space, $\Gamma=1$, and $\theta$ and is the angle between two 4d normals of the two adjacent polyhedra (the isotropy requires that this is the same for every couple of normals) and $K$ is proportional to the time derivative of the scale factor.

With these assumptions, any homogeneous isotropic coherent state on any regular graph is described by a single complex variable $z$, whose imaginary part is proportional to the area of each regular face of the cellular decomposition (and it can be put in correspondence with the total volume) and whose real part is related to the extrinsic curvature \cite{Freidel:2010aq}.  I denote 
this state as $\psi_{H_\ell(z)}$, and  the state on two copies of the regular graph, 
obtained tensoring the initial and a final homogeneous isotropic states, as $\psi_{H_\ell(z,z')}=\psi_{H_\ell(z)}\otimes \psi_{H_\ell(z')}$.

The classical Hamilton function of a homogeneous isotropic cosmology is the difference between two boundary terms. With the cosmological constant $\Lambda$ it gives
\be 
\!\!\!\!\!\!\!
S_H= 
\int {\mathrm{d}}t\, (a\dot a^2 + \frac\Lambda3 a^3)\Big|_{\dot a =\pm \sqrt{\frac\Lambda3} a}= 
\frac23 \sqrt{\frac\Lambda3} (a^3_{fin}-a^3_{in}).
\ee
where  $a$ is the scale factor and $\dot a$ its time derivative.
 Therefore at the first order in $\hbar$ the quantum transition amplitude factorizes:
\be  W(a_{fin},a_{in})=e^{\frac i \hbar \,S_H(a_{fin},a_{in})}=W(a_{fin})\overline{W(a_{in})} ~.\ee
The same happens for the spinfoam amplitude
\be\label{factoriz}\!\!\!\!\!
\langle W |\psi_{H_\ell(z_{fin},z_{in})}\rangle=W(z_{fin},z_{in}) = W(z_{fin})\, \overline {W(z_{in})}~~
\ee
where
\be\label{Wz}
W(z)= \int
\ \prod_{n=1}^{N\!-\!1}\! \d G_n ~
\prod_{\ell=1}^ {L}\  
   P_t\big(H_{\ell}(z) \, , \,G_\ell\big)~.
\ee%
The integration is on the group elements $G_n \in SL(2,\C)$, one for each node $n$.  We are interested in this quantity in the limit in which the imaginary part of $z$ of large, namely in the large volume limit. 

Let us start by studying \eqref{pirippipipi} when the imaginary part of $z$ is large.  
In the trace there is
\be
D^{\scriptscriptstyle(j)}(e^{-iz\f{\sigma_3}2})=\sum_m e^{-iz m}\ |m\rangle\langle m|
~.
\ee
For 
Im\,$z \gg 1$ (large area) in this sum the term $m=j$ dominates, therefore
\be
D^{\scriptscriptstyle(j)}(e^{-iz\f{\sigma_3}2})\approx e^{-iz j}\ |j \rangle\langle j|
\ee
where $|j\rangle$ is the the eigenstate of $L_3$ with maximum eigenvalue $m=j$ in the representation $j$.
Inserting this result into  \eqref{pirippipipi} and \eqref{Wz}  I obtain
\begin{multline}
\label{Wtostart}
W(z)= \int
\ 
\prod_{n=1}^{N\!-\!1}\! \d G_n ~
\prod_{\ell=1}^ {L}\  
\sum_{j_\ell} {\scriptstyle (2j_\ell+1)} e^{\scriptscriptstyle-2t\hbar j_\ell(j_\ell+1)
{ - i z_\ell j_\ell} }\\ \ \ \ 
\times 
\langle j_\ell|
{ D^{\scriptscriptstyle(j_\ell)}(R^{-1}_{\vec n_t})}\,
Y^\dagger  D^{\scriptscriptstyle(\gamma j_\ell, j_\ell)}\!(G_\ell) Y\!
{ D^{\scriptscriptstyle(j_\ell)}(R_{\vec n_s})}
 |j_\ell\rangle\,.
\end{multline}
The action of the matrix $D^{\scriptscriptstyle(j_{\!\ell})}\!(R_{\vec n_n})$ on the highest weights
states is precisely the definition of the coherent states $|\vec n\rangle$, so I can write
\be
\label{Wtostart}
W(z)&=& \int
\ 
\prod_{n=1}^{N\!-\!1}\! \d G_n ~
\prod_{\ell=1}^ {L}\  
\sum_{j_\ell} {\scriptstyle (2j_\ell+1)} e^{\scriptscriptstyle-2t\hbar j_\ell(j_\ell+1)
{ - i z_\ell j_\ell} }\nn
&&\ \ \ \times \ \
  \bra{\vec n_{t(\ell)}}\,
\,Y^\dagger  D^{\scriptscriptstyle(\gamma j_{\!\ell}, j_{\!\ell})}\!(G_\ell) Y\,
\ket{\vec n_{s(\ell)}}~.
\ee

I can now study the  $SL(2,\C)$ integral in \eqref{Wtostart} (without fixing the $j$). Let us rewrite the previous expression as
\begin{multline}
\label{Wsplit}
W(z)=
\sum_{\{j_\ell\}}\  \prod_{\ell=1}^ {L}  \!{\scriptstyle (2j_\ell+1)} \, e^{\scriptscriptstyle-2t\hbar j_\ell(j_\ell+1)
{ - i zj_\ell} }
\\ \times~
 \int \  \prod_{n=1}^{N\!-\!1}\! \d G_n 
\prod_{\ell=1}^ {L}
  \bra{\vec n_{t(\ell)}}\,
\,Y^\dagger  D^{\scriptscriptstyle(\gamma j_{\!\ell}, j_{\!\ell})}\!(G_\ell) Y\,
\ket{\vec n_{s(\ell)}}~.
\end{multline}
Since the gaussian sums in the first line peak the $j_\ell$'s over large values, the integral in the second line
can be computed in the large spin regime, where it can be evaluated using saddle point methods.  The computation of the integral 
\be \!\!\!\!\!\!\!\!\! \label{integrale}
 \int
 \  \prod_{n=1}^{N\!-\!1}\! \d G_n 
\prod_{\ell=1}^ {L} \ 
 \bra{n_{s(\ell)}}\,
\,Y^\dagger  D^{\scriptscriptstyle(\gamma j_{\!\ell}, j_{\!\ell})}\!(G_\ell) Y\,
\ket{n_{t(\ell)}}
\ee
is simplified in a spinor base, as the one introduced in \cite{Bianchi:2011ta} and gives  %
\be\!\!\!\!
\mathrm{H}\ 
 \prod_{\ell=1}^ {L}
e^{
-\frac12 i{ j_\ell}\,\theta
}
\ee
%
where H is the Hessian of the logarithm of the integrand in \eqref{integrale} \cite{Bianchi:2011ta} and $\theta$
is a constant determined by the normals on the faces: it is the \emph{intrinsic} curvature
on the faces, coming from the spin connection in the Ashtekar connection.
I can define a new variable $\tilde z:= z-\theta$, so that the real part of $\tilde z$ is exactly the extrinsic curvature. Using this, the ampliude becomes
\be\label{labello}
W(z)=
\sum_{\{j\}} \mathrm{H}\  \prod_{\ell=1}^ {L}  \!{\scriptstyle (2j_\ell+1)} \,
e^{\scriptscriptstyle-2t\hbar j_\ell(j_\ell+1)    - i \tilde z j_\ell} ~.
\ee
Since the imaginary part of $\tilde z$ is large, we can approximate the sum that appears in the amplitude with a Gaussian integral. I call $j_o$ the peak value of $j_\ell$, which is the same for all $\ell$.   
Following the same steps as \cite{Bianchi:2010zs,Bianchi:2011ym}, we can then rewrite the amplitude \eqref{labello}  as
\be
W(z)=
 \mathrm{H}\
\left( \sum_{j}  \!{\scriptstyle (2j+1)} \, 
e^{\scriptscriptstyle-2t\hbar j(j+1)    - i \tilde z j} 
\right)^L
\label{www}
\ee
where $H$, which is polynomial in $j$, is now taken 
at the stationary point $j_o$. Here the Hessian give a contribution $N_{\Gamma}$ that depends
 on the graph $\Gamma$ trough its numbers of links $L$ and nodes $N$,  and a characteristic term $j_o^{-3}$ that is independent of the graph. This is norm squared of the Livine-Speziale coherent regular cell of size $j_o$ \cite{Livine:2007vk} (recently calculated in the Lorentzian \cite{Bianchi:2011ta}). Notice that since I have fixed the normals $\vec n_n$, degenerate contributions are not allowed (these being present, I would have had further terms $\sim j_o^{-1}$).

The value of $j_o$ 
is determined by the vanishing of the real part of the exponent in  \eqref{www}.
This gives a condition on the imaginary part of $\tilde z$ (associated to the area). When this is large ($ j\gg1$), I have \be\label{Re} j_o\sim \textrm{Im}\,\tilde z/4t\hbar ~.\ee
The imaginary part of \eqref{www} is a phase that suppress the amplitude everywhere but where the argument is zero or a multiple of $2\pi$. This gives the condition
\ba\label{Im1}
&\textrm{Re}\, \tilde z = 0~,&
\ea
Using \eqref{Im2}, this is  
\ba\label{Im}
\theta\left(\gamma K+1
\right)-
\theta=0.
\ea
Without a source (matter or the cosmological constant) this implies $K=0$, namely $\dot a=0$, which is the only solution of the Friedmann equation in this case.

The final expression of the amplitude is
\be
W( z)= \left(\sqrt{{\f\pi t}}\ e^{-\f{\tilde z^2}{8t\hbar}}\ 2j_o  \right)^{\!\!L}\, \f{N_\Gamma}{j_o^{3}} \, \ee
%
so that, using this and \eqref{Re}, I conclude
\be\label{amplifon}
W( z)= N z^{L-3} e^{-\f{L}{2t\hbar} z^2}
\ee
where $N=(\fracs{4\pi}{t})^{\!L/2}\,(\f{-i}{4t\hbar} )^{\!L-3}\, N_{\Gamma}$
. Finally, inserting into \eqref{factoriz} I have
\be
W( z_\ini, z_\fin)= N^2\  ( z_\ini\,  z_\fin)^{\!L-3}\ e^{-\f{L}{2t\hbar}( z_\ini^2+ z^2_\fin)}~.
\label{eccola}
\ee

This is the transition amplitude between two cosmological homogeneous isotropic coherent states, with an arbitrary number of nodes $N$ and a number $L$ of links such that the graph is regular (i.e. every node has the same valency).

\section{Cosmological constant \\ and Friedmann equation}\label{secos}

It is useful to consider a modification of the transition amplitude in order to compare our result in the semiclassical limit beyond Minkowski space, which is the only solution in the absence of matter and cosmological constant. Following \cite{Bianchi:2011ym}, let us add a cosmological constant term  in \eqref{zeta} as follows 
\be
   Z_{\cal C}=\sum_{j_f,\vol_e} \prod_f (2j+1){~  \prod_e e^{i\lambda \vol_e \ } }\prod_v A_v(j_f,\vol_e). 
\label{modificazione}
\ee
where 
$\lambda$ is a constant\footnote{%
One can equivalently introduce an effective matter by replacing $\lambda$ with a density $\rho$. This will be studied elsewhere. 
} that yelds the cosmological constant $\Lambda$ and
$\vol_e$ is the volume associated to an edge: in presence of homogeneity and isotropy, all the cells are the same and I can write $\vol_e$ as the volume $\vol_o$ of a regular cell with faces having unit area, times $j^{\frac32}$.

The transition amplitude \eqref{www} becomes
\be
\!W_v(z)\!=\!
\sum_{j} \prod_{\ell=1}^ {L}  \!{\scriptstyle (2j_\ell+1)} \, \mathrm{H}\ 
e^{\scriptscriptstyle-2t\hbar j_\ell(j_\ell+1)    - i \tilde z j_\ell} 
e^{-i\lambda \vol_o j^{\frac32}
}
\label{cosmowww}
\ee
I expand around  $j_0$  so that the new term is 
\be
     i\lambda \vol_o j^{\frac32}\sim  i\lambda \vol_o j_o^{\frac32}+
     \fracs32 i\lambda \vol_o j_o^{\frac12}\delta j. 
\ee
The first term is a constant that can be reabsorbed in the normalization and the second contributes to
the phase such that the condition \eqref{Im1} becomes
\be
\textrm{Re}\, \tilde z = \fracs32
 \lambda \vol_o j^{\frac12}~.
 \label{zero}
\ee
At the stationary point condition \eqref{Re} holds so I obtain
\be\label{Rel}
\textrm{  Re}\,\tilde z = \fracs32\lambda \vol_o j_o^{\frac12} =  \fracs32\lambda \vol_o \sqrt{{ \textrm{  Im}\,\tilde z}/{4t\hbar}} .  
\ee 
This expression yields the Friedmann equation:
recall that $\textrm{  Re}\,\tilde z \sim \dot a$ and $\textrm{  Im}\,\tilde z \sim a^2$ so that,
squaring the previous equation, I obtain 
\be\label{fried}
  \left(\frac{\dot{a}}{a}\right)^2=\frac{\Lambda}{3}, 
        \label{ds}
\ee
where $\Lambda= 27\,\lambda^2\vol_o^2/16\,t\hbar$. 
The same result can be obtained by a different technique: the transition amplitude results to be annihilated by a Hamiltonian constraint. In the classical limit, this is 
\ba
    ( \tilde z+ \fracs32  \lambda \vol_o j_o^{\frac12})^2+\overline{( \tilde z+ \fracs32  \lambda \vol_o j_o^{\frac12})^2}=0 
\hskip15mm    \\
\mbox{that gives}\hskip13mm
i4\, \textrm{Im}\, \tilde z\ (\textrm{Re}\, \tilde z + \fracs32 \lambda \vol_o j_o^{\frac12} )=0~.
\hskip13mm
\ea
that is equivalent to \eqref{zero}.

Notice that 
I don't obtain the curvature term $k/a^2$ in the full Friedmann equation 
\be
  \left(\frac{\dot{a}}{a}\right)^2=\frac{\Lambda}{3} - \frac k{a^2}~.
        \label{ds}
\ee
This is because of the approximation taken in the evaluation of the gaussian sum. Since we ask for large $j$, namely for a large distance regime, the curvature term is neglected being a higher order in $1/j$ \cite{Magliaro:2011uq}. Finding a way to relax this approximation is an urgent issue in spinfoam cosmology: the higher order in $1/j$  would in fact provide us also the first quantum corrections.

The volume $\vol_o$ depends on the graph used. On the other hand, such a  cosmological-constant term has been introduced as an edge amplitude. This edge amplitude can be viewed as a redefinition of the vertex. Possible normalization ambiguities, coming from the introduction of this term, can therefore be absorbed in the vertex amplitude \cite{Magliaro:2010vn}.%

The transition amplitudes presented in this work are in fact  not normalized. The arbitrary normalization of the vertex amplitude is fixed by cylindrical consistency \cite{Magliaro:2010vn}.
Notice that the presence of many nodes enters only in the term $N$ in  \eqref{eccola}, and it can be counterbalanced by normalizing appropriately the amplitude. 

\vskip1em
The result of this calculation is that in the limit for large $j$, the support of the transition amplitude, obtained trough the conditions on the real and the imaginary part of $\tilde z$ that yields the Friedmann equation, is not sensitive to the number of links or the number of nodes of the graph used.




\section{Examples}\label{examples}

Let us illustrate some concrete 	\emph{regular} graphs for which the results above apply. I illustrate two concrete examples of boundary graphs: a graph with 2 nodes and many links, that has been used as a base for cosmological model also in the $U(N)$ framework \cite{Borja:2011di}, and the 4-simplex formed by 5 nodes completely connected, that is the most exploited graph in the spinfoam calculations.
\vskip1em

{\noindent\bf Many-links dipole}
\vspace{-1em}
\begin{figure}[h]
\begin{center}
\begin{picture}(50,50)\label{figdipole}
\setlength{\unitlength}{0.05in}
\put(00,00) {\circle*{1,6}} 
\put(20,00) {\circle*{1,6}} 
\put(-5,-1) {$G_1$} 
\put(22,-1) {$G_2$} 
\qbezier(00,00) (10,21) (20,00)
\qbezier(00,00) (10,10) (20,00)
\qbezier(00,00) (10,06) (20,00)
\qbezier(00,00) (10,02) (20,00)
\qbezier(00,00) (10,-21) (20,00)
\qbezier(00,00) (10,-10) (20,00)
\qbezier(00,00) (10,-06) (20,00)
\qbezier(00,00) (10,-02) (20,00)
\put(10,08) {:}\put(10,07) {.}   
\put(10,-8) {:} \put(10,-9) {.}    
{\color{celeste} \put(12,11) {$j_L$}  \put(12,-8.5) {$j_1$} }
\end{picture} \end{center}
\vspace{3em}
\caption{The graph $\Gamma_2$: a ``dipole" with $L$ links.}
\end{figure}
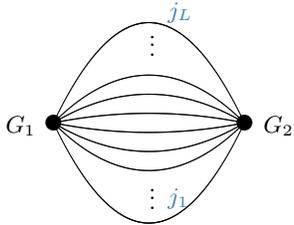

A first generalization is given by adding more links to a dipole graph, as in 
the figure above.  The presence of 
%
 only 2 nodes greatly simplified the calculation. In particular, it simplifies the integration on the group elements $G_n$ since it is possible to define an unique integration variable $G=G_1G_2^{-1}$. The vertex amplitude \eqref{Wz} becomes
\be\label{WzL}
W_v(z)= \int_{SL(2,\C)}\!\! \d G  ~\ 
   P_t(H_{z} \, , \,G)^L
\ee
with $P_t(H_{z}  , G)$ as in \eqref{pirippipipi}. Let us perform first the integration in $G$ by the saddle point approximation around $j_o$, 
obtaining
\begin{eqnarray}
W(z)&=& \left(
\sum_{j} { (2j+1)} \ e^{-2t\hbar j(j+1)- i \tilde zj      
}\right)^{\!\!\! L} 
\frac{N_{\Gamma_2}}{j_o^3}
\label{Wdipole}
\end{eqnarray}
where $N_{\Gamma_2}$ is a constant that depends on the number of links $L$ and can be absorbed in the normalization. 
Notice that in this case the 4-dimensional normals between the polyhedra at each nodes have to be parallel and therefore $\theta=\pi$. 

I study the support of the transition amplitude, getting the condition on the real and the imaginary part of $z$ \eqref{Im} and \eqref{Re}, or \eqref{Rel} with the cosmological constant.
Notice that these conditions do not depend on $L$.
Therefore the support of the amplitude doesn't depend of the number of links in the dipole graph. I conclude that the vertex amplitude from the EPR/FK/KKL model, in the homogeneous and isotropic case,  bears the Friedmann dynamics independently of the number of links in the chosen graph.

The final result by performing the gaussian integral is given by \eqref{amplifon}
where now $N_\Gamma = N_{\Delta}$.
\vskip5pt
The phase space and the canonical dynamics associeted to this graph has been studied in details in the the $U(N)$/spinor framework \cite{Girelli:2005ii,Freidel:2010tt,Borja:2010gn,Borja:2010rc,Livine:2011gp}. It would be interesting to compare the definition of the transition amplitude in terms of the spinors 
 with \eqref{Wdipole}.

\vskip1em
{\noindent \bf The 4-simplex graph}

The 3-sphere is a natural geometry for modeling our universe \cite{Einstein:1917ce,Peterson:1979fk}
and the simplest non-degenerate triangulation is given by the complete graph on five nodes $\Gamma^5$. The coherent states for this graph has been studied in detail in \cite{Magliaro:2011kx}. 
I can apply explicitly \eqref{www} and \eqref{eccola} to obtain the transition amplitude between two states that live on such a graph.

%

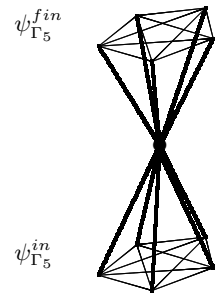
\begin{figure}[h]
\begin{center}
\begin{picture}(20,115)
\setlength{\unitlength}{0.04in}
\put(-10,35) {$\psi_{\Gamma_5}^{fin}\  $}
\qbezier(01,33)(06,37)(06,37)
\qbezier(01,33)(15,38)(15,38)
\qbezier(01,33)(16,34)(16,34)
\qbezier(01,33)(08,31)(08,31)
\qbezier(06,37)(06,37)(15,38)
\qbezier(06,37)(06,37)(16,34)
\qbezier(06,37)(06,37)(08,31)
\qbezier(15,38)(15,38)(16,34)
\qbezier(16,34)(16,34)(08,31)
\qbezier(15,38)(08,31)(08,31)
\put(-10,05) {$\psi_{\Gamma_5}^{in}\  $}
\qbezier(01,03)(06,07)(06,07)
\qbezier(01,03)(15,08)(15,08)
\qbezier(01,03)(16,04)(16,04)
\qbezier(01,03)(08,01)(08,01)
\qbezier(06,07)(06,07)(15,08)
\qbezier(06,07)(06,07)(16,04)
\qbezier(06,07)(06,07)(08,01)
\qbezier(15,08)(15,08)(16,04)
\qbezier(16,04)(16,04)(08,01)
\qbezier(15,08)(08,01)(08,01)
\linethickness{0.4mm}
\put(09,20) {\circle*{1,7}} 
\qbezier(09,20)(09,20)(06,07)
\qbezier(09,20)(09,20)(15,08)
\qbezier(09,19)(09,19)(16,04)
\qbezier(09,20)(09,20)(08,01)
\qbezier(09,20)(09,20)(01,03)
\qbezier(09,20)(09,20)(06,37)
\qbezier(09,20)(09,20)(15,38)
\qbezier(09,20)(09,20)(16,34)
\qbezier(09,20)(09,20)(08,31)
\qbezier(09,20)(09,20)(01,33)
\end{picture} 
\end{center} 
\caption{Transition amplitude between two states defined on $\Gamma_5$ graphs.}
\end{figure}

%

In this case the transition amplitude for one connected component is as in \eqref{WzL}, with $L=10$ and a factor $N_{\Gamma_5}$ that carries the information about the number of nodes in the graph.
The value of $\theta$ is well-know and it is equal to $\arccos(-1/4)$.

This transition is a natural candidate to further studies in spinfoam cosmology, such as cosmological perturbations theory.


\section{Conclusions}
 
 I have computed the spinfoam transition amplitude for states peaked on a homogenous and isotropic geometry, introducing two improvements with respect to the previous works: the amplitude is now Lorentzian and it has been generalized for every regular graph, with an arbitrary number of links and of nodes.
 
 The oscillating phases of the amplitude suppresses the sum everywhere but where the imaginary part of the exponent vanishes: this gives a condition on the real part of $z$ (i.e. on the area).
The gaussian sum is peaked on the maximum of the real value of the exponent, and this give a condition on the imaginary part of $z$ (i.e. on the extrinsic curvature). These two conditions together yield the Friedmann equation.
 
In particular, these conditions holds independently of the number of nodes or the number of links that are present in the graph: this is the main result presented in the paper.
This shows that the results obtained in the previous works are robust with respect to different choices of graph on which the boundary states are defined. 
 
I have evaluated the amplitude before performing the gaussian integral in $j$: this allows to study its periodicity. The gaussian integral is usually performed in the large $j$ limit, in a way that washes away most of the quantum effects such as the periodicity of $W(z)$  in the real part of $z$ (associated to the extrinsic curvature). This is particularly interesting in relation with the $\bar\mu$-scheme that is used in loop quantum cosmology \cite{Ashtekar:2006wn}. The difference between the ``old'' scheme and the ``new'' $\bar\mu$ quantization scheme can be looked from the perspective of which fundamental variable emerges as periodic after the quantization: for the former it is the time derivative of the scale factor $\dot a$, for the latter it is the Hubble rate $\dot a/a$. Different quantization schemes have been proposed in loop quantum cosmology, but the $\bar\mu$ one seems to give the most robust predictions \cite{Corichi:2008fu}.
It is therefore highly desirable to see a convergence of canonical and covariant formalism by finding a periodicity in the Hubble rate in the amplitude. The present formulation of the spinfoam amplitude seems to give instead a periodicity in $\dot a$. This does appear to affect the classical large-distance behavior in this context, but it questions the regime of validity of the approximations taken, when quantum corrections are concerned.  In which regime does the truncation taken correctly describes the quantum corrections, and where should it be modified in order to match the $\bar\mu$-scheme approximation? Work is in progress to study these questions, for instance considering averaging over many nodes or graph-changing transitions.

An important open issue is to compute the corrections that appear when considering more than one vertex in the spinfoam. We are not interested in a mere sequence of edges and vertex, because it has to be equivalent to a single vertex \cite{Bahr:2010bs}. We would like to have instead spinfoam faces spanning from the initial to the final states and carrying the correlations between the two states 
 (see FIG.\,4). For consistency, these higher order spinfoams should not affect the the large $j$ limit of the amplitude.
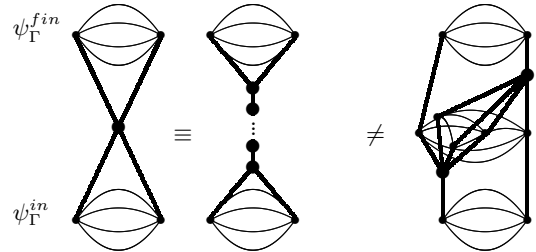
\begin{figure}[t]
\label{tre}
  \vskip10mm
  \begin{center}
\begin{picture}(20,40)
\put(-20,40) {$\psi_\Gamma^{fin}$} 
\put(-20,-30) {$\psi_\Gamma^{in}$} 
\put(04,40) {\circle*{3}} 
\put(36,40) {\circle*{3}}  
\qbezier(4,40)(20,63)(36,40)
\qbezier(4,40)(20,31)(36,40)
\qbezier(4,40)(20,49)(36,40)
\qbezier(4,40)(20,17)(36,40)
\put(04,-30) {\circle*{3}} 
\put(36,-30) {\circle*{3}}  
\qbezier(4,-30)(20,-53)(36,-30)
\qbezier(4,-30)(20,-21)(36,-30)
\qbezier(4,-30)(20,-39)(36,-30)
\qbezier(4,-30)(20,-07)(36,-30)
\linethickness{0.4mm}
\put(20,5) {\circle*{5}} 
\qbezier(4,39)(04,39)(36,-29)
\qbezier(4,-29)(36,39)(36,39)
\end{picture} 
\hskip3em
\begin{picture}(20,40)
\put(04,40) {\circle*{3}} 
\put(36,40) {\circle*{3}}  
\qbezier(4,40)(20,63)(36,40)
\qbezier(4,40)(20,31)(36,40)
\qbezier(4,40)(20,49)(36,40)
\qbezier(4,40)(20,17)(36,40)
\put(04,-30) {\circle*{3}} 
\put(36,-30) {\circle*{3}}  
\qbezier(4,-30)(20,-53)(36,-30)
\qbezier(4,-30)(20,-21)(36,-30)
\qbezier(4,-30)(20,-39)(36,-30)
\qbezier(4,-30)(20,-07)(36,-30)
\linethickness{0.4mm}
\put(20,-2) {\circle*{5}} 
\put(20,20) {\circle*{5}} 
\put(20,12) {\circle*{5}} 
\put(20,-10) {\circle*{5}} 
\put(19,4) {.}
\put(19,6) {.}
 \put(19,2) {.}
\qbezier(20,-10)(20,0)(20,0)
\qbezier(20,10)(20,20)(20,20)
\qbezier(20,-10)(20,-10)(36,-29)
\qbezier(20,-10)(20,-10)(04,-29)
\qbezier(20,20)(20,20)(36,39)
\qbezier(20,20)(20,20)(04,39)
\put(-10,0) {$\equiv$}
\end{picture} 
\hskip7em
\begin{picture}(20,40)
\put(-25,0) {$\neq$}
\put(04,40) {\circle*{3}} 
\put(36,40) {\circle*{3}}  
\qbezier(4,40)(20,63)(36,40)
\qbezier(4,40)(20,31)(36,40)
\qbezier(4,40)(20,49)(36,40)
\qbezier(4,40)(20,17)(36,40)
\put(04,-30) {\circle*{3}} 
\put(36,-30) {\circle*{3}}  
\qbezier(4,-30)(20,-53)(36,-30)
\qbezier(4,-30)(20,-21)(36,-30)
\qbezier(4,-30)(20,-39)(36,-30)
\qbezier(4,-30)(20,-07)(36,-30)
\qbezier(-6,03)(14,20)(36,03)
\qbezier(20,03)(12,12)(02,09)
\qbezier(-4,02)(14,-10)(36,03)
\qbezier(-5,02)(18,-19)(36,03)
\qbezier  (8,-3)  (18,01) (20,03) 
\qbezier  (18,01) (25,07) (36,03)
\qbezier  (8,-2)  (06,10)(02,09)
\qbezier  (-5,03)(5,08)(20,03)
\linethickness{0.4mm}
\put(4,-12) {\circle*{5}} 
\put(36,25) {\circle*{5}} 
\qbezier(36,-30)(36,10)(36,40)
\qbezier(4,-29)(4,-10)(4,-10) 
\qbezier(-5,03) (4,-12) (4,-12) 
\qbezier(21,03) (4,-12) (4,-12) 
\qbezier  (8,-3)  (4,-12) (4,-12) 
\qbezier(02,09)(36,25)(36,25)
\qbezier(21,03) (36,24)(36,24)
\qbezier  (8,-2)  (36,25)(36,25)
\qbezier(02,09) (4,-12) (4,-12) 
\qbezier(-5,03) (4,40) (4,40) 
\put(-5,03) {\circle*{3}}  
\put(20,03) {\circle*{3}}  
\put(36,03) {\circle*{3}}  
\put(02,09) {\circle*{3}}  
\put(8,-2) {\circle*{3}} 
\end{picture} 
\end{center} 
\vskip15mm
\caption{The two images on the left represent transition amplitudes at the first order in the vertex expansion. The third graph is an example of higher order transition.}
\end{figure}

\newpage
Then, it would be of great importance to explore quantum effects by going  beyond the  large-$j$ regime. In the low-$j$ regime, we expect the dynamics to depends on the graph. Finally, we would like to explore different schemes to obtain the semiclassical limit, such as 
the double scaling limit $\gamma\rightarrow 0$, $j\rightarrow\infty$ where the physical area is kept constant \cite{Magliaro:2011fk,Bianchi:2011ta}.

 \vskip5mm
 \noindent
{\bf Acknowledgments} ~
A warm thank to Carlo Rovelli and Eugenio Bianchi for several valuable discussions, and to Simone Speziale and Ed Wilson-Ewing for useful comments on the draft of this paper.

 \bibliographystyle{apsrev4-1}   
\bibliography{bib,BiblioCarlo}
 \end{document}